\documentclass{vgtc}                          

\graphicspath{{figures/}{pictures/}{images/}{./}} 

\usepackage{times}                     

\usepackage{tabu}                      
\usepackage{booktabs}                  
\usepackage{lipsum}                    
\usepackage{mwe}                       
\usepackage{amsmath}
\usepackage{color}
\usepackage{balance}
\usepackage{mathptmx}                  

\onlineid{0}

\vgtccategory{Research}

\vgtcinsertpkg

\ieeedoi{xx.xxxx/TVCG.201x.xxxxxxx}

\title{CLAd-VR: Cognitive Load-based Adaptive Training for Machining
Tasks in Virtual Reality}

\author{Bhavya Matam\thanks{e-mail: bm3792@nyu.edu}\\ %
        \scriptsize New York University %
\and Adamay Mann\thanks{e-mail: am14579@nyu.edu}\\ %
     \scriptsize New York University %
\and Kachina Studer\thanks{e-mail: xkachina@mit.educh}\\ %
     \scriptsize Massachusetts Institute of Technology %
\and Christian Gabbianelli\thanks{e-mail: ctgabb@mit.edu}\\ %
     \scriptsize Massachusetts Institute of Technology %
\and Sonia Castelo\thanks{e-mail: s.castelo@nyu.edu}\\ %
     \scriptsize New York University. %
\and John Liu\thanks{e-mail: johnhliu@mit.edu}\\ %
     \scriptsize Massachusetts Institute of Technology %
\and Claudio Silva\thanks{e-mail: csilva@nyu.edu}\\ %
     \scriptsize New York University %
\and Dishita Turakhia\thanks{e-mail: turakhia.d@nyu.edu}\\ %
     \parbox{1.4in}{\scriptsize \centering New York University}}

\teaser{
  \centering
  \includegraphics[width=\linewidth]{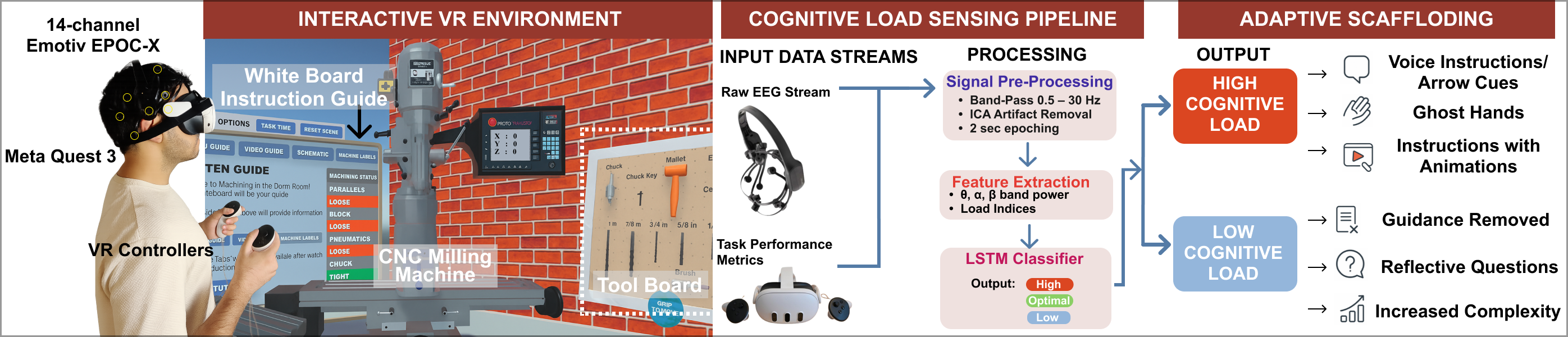}
  \caption{CLAd-VR is a cognitive load–based adaptive VR training system for precision machining tasks, consisting of: A) an interactive multimodal training module in VR; B) a cognitive load sensing pipeline, and C) an adaptive scaffolding engine.}
  \label{fig:teaser}
}

\abstract{

With the growing need to effectively support workforce upskilling in the manufacturing sector, virtual reality (VR) is gaining popularity as a scalable training solution. However, most current systems are designed as static, step-by-step tutorials and do not adapt to a learner's needs or cognitive load—a critical factor in learning and long-term retention. We address this limitation with CLAd-VR, an adaptive VR training system that integrates real-time EEG-based sensing to measure the learner's cognitive load and adapt instruction accordingly, specifically for domain-specific tasks in manufacturing. The system features a VR training module for a precision drilling task, designed with multimodal instructional elements including animations, text, and video. Our cognitive load sensing pipeline uses a wearable EEG device to capture the trainee’s neural activity, which is processed through an LSTM model to classify their cognitive load as low, optimal, or high in real time. Based on these classifications, the system dynamically adjusts task difficulty and delivers adaptive guidance using voice guidance, visual cues, or ghost-hand animations. This paper introduces CLAd-VR system's architecture, including the EEG sensing hardware, real-time inference model, and adaptive VR interface.

} 


\nocopyrightspace

\begin{document}



\maketitle
\section{Introduction}
The manufacturing sector faces a growing skills gap with up to 2 million U.S. jobs projected to go unfilled by 2030~\cite{wellener2018skills}. Virtual Reality (VR) training has emerged as a promising scalable alternative~\cite{radianti2020systematic} to traditional resource-intensive training methods that are falling short in narrowing the skills gap~\cite{makransky2019adding}. VR immersive environments can simulate real-world workflows, allowing trainees to safely practice procedures with repeatability~\cite{alshaer2020observer} and thus be useful for training. However, current VR-based training systems are typically designed as static on pre-scripted task sequences and feedback~\cite{rahimi2025generative}, offering the same instructional path to all users regardless of the individual trainee’s needs and cognitive load—which is critical for long-term skill learning~\cite{sweller1988cognitive}. Low cognitive load during training may lead to disengagement~\cite{paas2003cognitive}, while excessive load can overwhelm and hinder retention~\cite{gkintoni2025challenging}. Therefore, adaptive training systems based on the real-time estimation of cognitive load are essential to support optimal training~\cite{gerjets2014cognitive}.

This paper presents the design of our work-in-progress system,  CLAd-VR - a cognitive-load-based adaptive VR system for manufacturing training. 
The system has three core components: a VR training module, a cognitive load sensing pipeline, and an adaptive feedback engine. The training module is a Unity-based environment for precision CNC drilling task training implemented in a Meta Quest 3 headset with multimodal instructional content like animations, voice guidance, and visual overlays. To assess cognitive load in real time, we use a wearable EEG headset (Emotiv EPOC X), which sends data to a lightweight LSTM model trained to classify neural activity into low, optimal, or high cognitive states. These predictions are informed by both EEG signals and task performance metrics. Based on the inferred load, the CLAd-VR system adjusts the task complexity and scaffolding to maintain optimal engagement according to our adaptation strategies. 

In summary, this paper presents CLAd-VR (see Fig. \ref{fig:teaser}), which has: a VR-based interactive training environment with multimodal instructional content, (2) a machine-learning-based cognitive load sensing and classification pipeline, and (3) an adaptive logic engine that modulates guidance and task difficulty during training.

\section{Related Work}

In this section, we first provide the theoretical background on Cognitive Load, discuss existing methods to assess it, and highlight the gap in related work in cognitive-load-aware adaptive systems for skill learning. 
\subsection{Cognitive Load Theory}
We ground our work in Cognitive Load Theory (CLT), which provides a robust theoretical framework to understand how humans process information and learn~\cite{sweller1988cognitive}. CLT posits that learners experience three types of cognitive load: {(i) intrinsic load}, which comes from how complex the learning material or task is, {(ii) extraneous load}, which is caused by poorly designed instructions or irrelevant information, and {(iii) germane load}, which is the effort the learners put into making sense of the material, connecting new ideas to what they already know, and building mental models. CLT emphasizes that the design of educational technologies should aim to minimize extraneous cognitive load thereby preserving the learner's working memory for investing in germane load. This is because our working memory, which holds and processes new information in the short-term, has limited capacity~\cite{Sweller2011-aa}. In the context of manufacturing and psychomotor skill training, high cognitive strain can hinder motor task execution, coordination, and accuracy~\cite{Jaquess2018-iz} while underload risks disengagement and lapses in attention~\cite{Langner2013-yj}. Therefore, for optimal skill acquisition and transfer, assessing and using cognitive load becomes essential to design effective adaptive learning environments in skill-intensive domains like manufacturing~\cite{Studer2024}. 

\subsection{Cognitive Load Assessment}

Cognitive load is typically assessed through subjective reports (like the {NASA Task Load Index (NASA-TLX)}~\cite{HART1988139} and CLT-based questionnaires~\cite{Leppink2013-sr}), performance metrics (such as task time and error rates~\cite{paas1994instructional}), and physiological sensing (like eye-tracking, heart rate variability (HRV), and electroencephalography (EEG)). Compared to commonly used subjective tests, physiological sensing offers a more reliable, real-time alternative. For example, EEG-based measures, such as increased frontal theta and reduced parietal alpha power, are widely used to detect cognitive workload~\cite{berka2007eeg, brouwer2012estimating}, and recent advances in machine learning have pushed the accuracy of EEG-based classification even higher. While significant progress has been made in EEG-based cognitive load detection and adaptive interfaces, most VR training systems for manufacturing remain static. They do not adapt to a learner's cognitive state, and existing cognitive load detection methods are often validated in simplified lab tasks, not complex, real-world training scenarios. Furthermore, integrating the necessary hardware, like the Emotiv EPOC X EEG headset with a Meta Quest 3, presents technical challenges that must be addressed to ensure low-latency, synchronized data streaming for real-time adaptation. Our CLAd-VR system addresses these gaps by providing a complete, integrated pipeline that combines real-time EEG-based cognitive load detection with adaptive VR training specifically designed for precision manufacturing tasks.


\subsection{Adaptive Immersive Training}
Designing instructions and scaffolding to optimize cognitive load (CL) is especially critical in VR, where immersion can unintentionally increase mental demand. Recent studies have shown that learners in immersive VR environments can experience higher CL, which impairs motor memory consolidation, task accuracy, and skill transfer~\cite{Juliano2022, Frederiksen2019}. This underscores the need for adaptive scaffolding that dynamically responds to learner CL. There are two key strategies for personalization in training are adapting task difficulty and adapting support or feedback. While step-by-step guidance systems help structure procedural learning, static instructions may disengage learners when cognitive load is too high or too low~\cite{turakhia2021can, turakhia2021adapt2learn}. In immersive AR/VR, task guidance has been shown to enhance efficiency and reduce cognitive load across domains like cooking, surgery, and maintenance~\cite{castelo_argus_2024, bichlmeier2007laparoscopic, palmarini2018systematic}. AR/VR also improves engagement and highlights task-relevant information when guidance is designed appropriately~\cite{nijholt2022towards}. Despite the benefits of adaptive systems, in the context of manufacturing, most VR training systems rely on static, heuristic-based guidance and lack the intelligence to adapt in real time to individual learner states. We address this gap by developing a cognitive load-aware adaptive VR training system for manufacturing skills, as detailed next.


\section{CLAd-VR System Architecture}
CLAd-VR's architecture has the following three components: 
(1) a VR-based training module with multimodal instructional content, (2) a cognitive load sensing and classification pipeline, and (3) an adaptive logic engine that modulates guidance and task difficulty.

\subsection{Interactive VR Training Environment}
CLAd-VR's training environment is built in Unity and experienced via a Meta Quest 3 headset. The training module simulates a CNC milling task workflow with using 2 modules and 8 steps. The design of CLAd-VR offers trainees the freedom to interact anytime with tools, parts, or switch between task training modules. The environment has integrated multimodal instructional cues—animations, voice guidance, text overlays, and prompts to support learning. These instructional modalities are scaffolded by default with video instructions, schematics, and written guides, but are later adapted based on the trainee's cognitive load. Trainees go through a step-by-step sequence of a drilling task on a 3-axis mill, including steps like part alignment, tool mounting, and controlled milling. As the trainees go through the steps, our system records their task performance with contextual metadata including step ID, completion time, and any procedural errors (e.g., tool collisions or skipped actions). This task-level data is then combined with the EEG sensor data stream to comprehensively understand the trainee's cognitive state in real-time during the training.

\subsection{Cognitive Load Measuring Pipeline}
CLAd-VR's cognitive load sensing and inference pipeline consists of two steps: (1) synchronization of EEG sensor data and task performance details, and (2) real-time inference of cognitive load using a learned classification model.

\begin{figure}[h!]
    \centering
    \includegraphics[width= \linewidth]{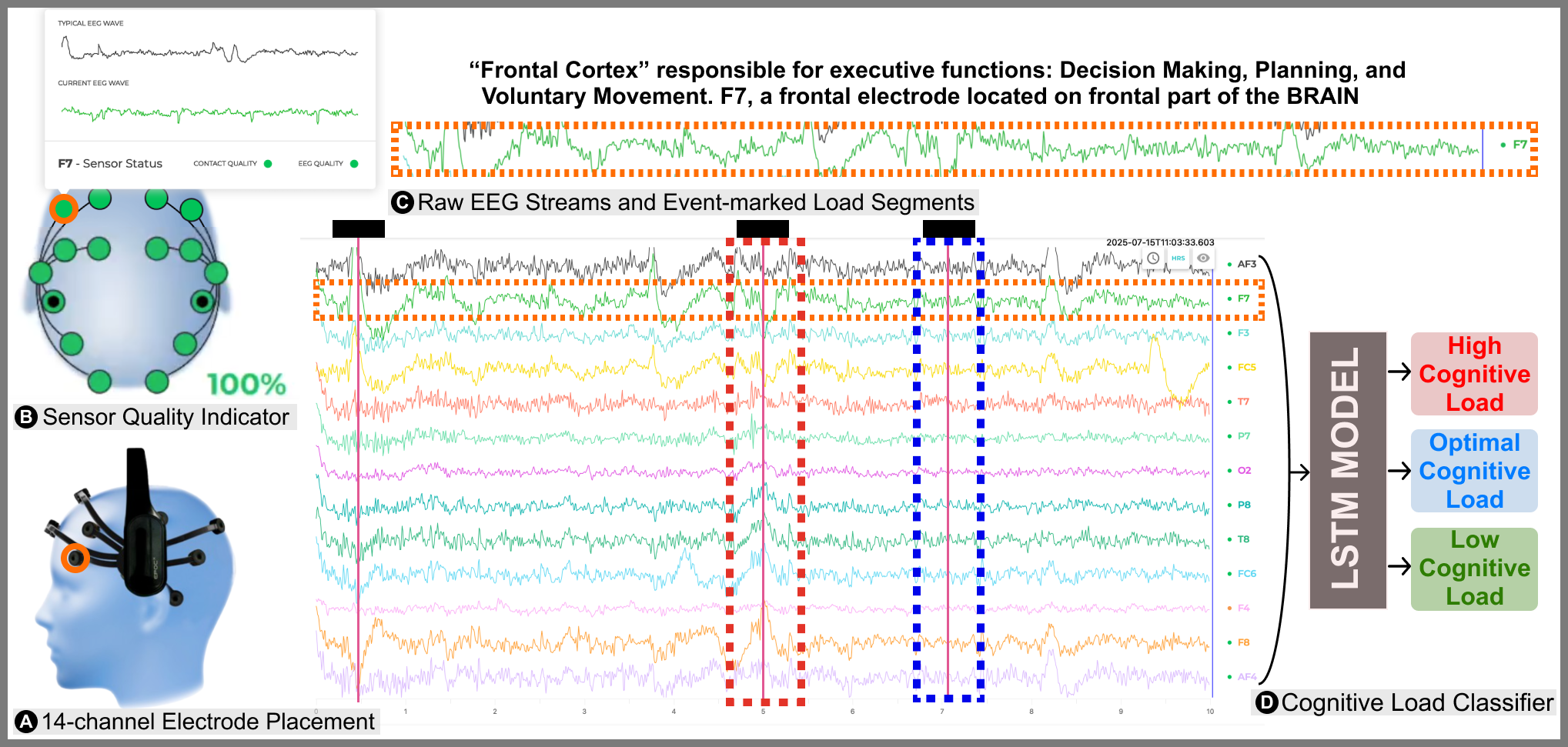}
    \caption{Real-time EEG data stream across 14 channels with instances showing high CL (in red) and optimal CL (in blue).}
    \label{fig:sampling}
    \vspace{-15 pt}
\end{figure}

\subsubsection{Synchronization of EEG Data and Task Performance}
During each CLAd-VR training session, the user wears an Emotiv EPOC X EEG headset~\cite{emotiv2021epocx} alongside a Meta Quest 3 VR headmounted device (HMD)~\cite{meta2023quest3}. This setup enables our system to capture synchronized neural and behavioral data in real time through two primary data streams: (i) EEG Signals: The EPOC X records brain activity from 14 scalp-mounted electrodes, arranged according to the standard 10–20 system~\cite{klem1999ten} (see Fig.~\ref{fig:sampling}). These signals are streamed at 128 Hz using the Lab Streaming Layer (LSL), allowing precise synchronization. From the raw EEG data, we extract frequency-domain features across theta (4–7 Hz), alpha (8–13 Hz), beta (14–30 Hz), and gamma (30–50 Hz) bands, which are the well-established markers for measuring cognitive load~\cite{klimesch1999eeg}. We also use features such as mean theta power and the theta-to-alpha ratio to study variations in CL. (ii) Task Performance Metrics: In parallel, the Unity-based VR environment continuously logs interaction data, including step IDs, task completion times, and error events. These logs are timestamped and aligned with EEG data via LSL, enabling multimodal, temporally precise analysis of how cognitive states evolve in relation to specific task actions. Together, these synchronized data streams form the foundation of our real-time CL classification pipeline.


\subsubsection{Calibration and LSTM Training}
To define cognitive load for every user, the system first runs a calibration phase. The phase uses an n-back test to create low and high mental workload~\cite{Jaeggi2010}. First we record 60 seconds of EEG data while the user rests to set a neural baseline. Next, the user does a simple 1-back task for two minutes. This creates a state of low cognitive load. Then, the user performs a harder 3-back task for two minutes to induce a high cognitive load. The system uses the EEG data gathered during these tasks to set the user's individual $T_{\text{low}}$ and $T_{\text{high}}$ thresholds respectively. This calibrated data helps train a Long Short-Term Memory (LSTM) neural network~\cite{Yang2020, Li2022}. Every 10 seconds, the system build a feature vector from the previous two seconds of data. This vector includes EEG-derived features (theta, alpha, and beta band power; the theta/alpha power ratio; and signal entropy) and behavioral metrics (task error count, step completion time, and current task difficulty level). The LSTM network has two hidden layers with 64 units each and uses a ReLU activation function. A Dropout Layer is included after each LSTM layer to help prevent the model from overfitting. The final output is a Dense layer with three units that uses a Softmax activation function. This function converts the model's output into a probability for each of the three class: low, optimal or high cognitive load. The model is trained using Adam optimizer and a Categorical Cross-Entropy loss function.

\subsubsection{Real-time Inference of Cognitive Load}
During the VR training session, the system processes the synchronized data streams in near real-time to infer the user's cognitive load. The feature vector, as defined in the previous section, is constructed and fed into the trained LSTM network. The LSTM architecture was chosen for its ability to model temporal dependencies in sequential physiological signals. The model outputs a continuous cognitive load score $L \in [0,1]$. This score is then compared to the user-specific thresholds established during calibration to infer one of three discrete states: (i) \textbf{Low Load} ($L \le T_{\text{low}}$) indicating low engagement or boredom, (ii) \textbf{Optimal Load} ($T_{\text{low}} < L < T_{\text{high}}$) indicating an effective zone of mental effort, or (iii) \textbf{High Load} ($L \ge T_{\text{high}}$) indicating overload, confusion, or frustration. Using these states, our system then adapts the training based on the strategies listed next. The full sensing-to-inference cycle maintains a latency under 100 milliseconds, enabling the system to respond quickly with adaptive guidance.

\subsection{Adaptation Strategies}
Next, our system uses the inferred cognitive load (CL) to adapt the scaffolding for the trainee in two ways: (i) adapting multimodal instructions and support and (ii) adjusting the training task difficulty. 

\begin{figure}[h!]
    \centering
    \includegraphics[width=\linewidth]{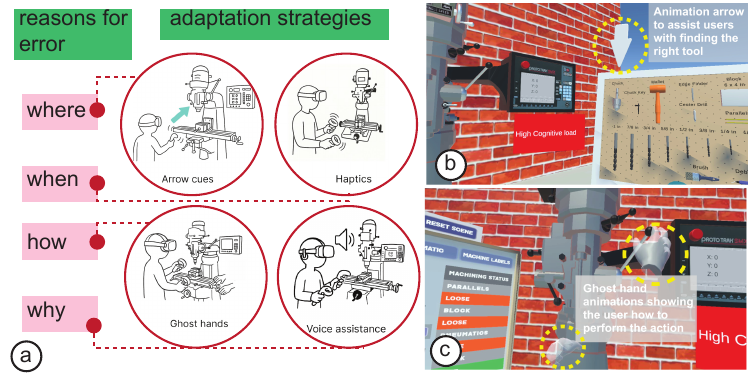}
    \caption{When the cognitive load is high, CLAd-VR adapts scaffolding to support the user using (a) Adaptation strategies based on the types of errors (b) example: arrow cues appear to guide the user’s actions; (c) example: ghost hands demonstrate the correct task steps to facilitate learning.}
    \label{adaptivestrategy}
    \vspace{-10 pt}
\end{figure}

When \textbf{high cognitive load} is detected, the system reinforces instructional scaffolding by simplifying interfaces, slowing task progression, and layering multimodal feedback. The strategies are selected according to the possible reasons for confusion (see Fig. \ref{adaptivestrategy}-a): “Where” is the spatial confusion (e.g., not knowing the location of a tool or machine part). The system deploys arrow-based cues highlighting the relevant component. “How” is the procedural confusion (e.g., the learner might not know how to place parallels in the vise). The system responds with ghost hand demonstrations that model the correct action. “Why” is the conceptual confusion (e.g., not understanding the role of parallels or the purpose of a tool) the system supplements with voice-based explanations and conceptual guidance. Additional support like haptic feedback is triggered through controller vibrations when the user holds an incorrect object, drawing immediate attention to the error. These interventions reduce ambiguity, provide just-in-time clarification, and keep trainees from becoming stuck or disengaged during critical phases of the machining workflow. For instance, as shown in Fig. \ref{adaptivestrategy}-b,c, arrow cues guide users toward the correct tool, while ghost hand animations demonstrate the required action.

When \textbf{low cognitive load} is detected, indicating that the trainee is under-challenged, the system reduces external guidance and increases task difficulty through three strategies: (i)~Error Injection Challenge where the system deliberately introduces misconfigurations (e.g., incorrect block placement) requiring the trainee to diagnose and correct the issue; (ii)~Adding Reflective Prompts where open-ended questions such as “Drill a hole at coordinates (3 cm, 4 cm)” require planning and multi-step execution are introduced; (iii)~Adding Time Pressure where the sub-tasks must be completed within specific time limits, simulating real-world machining urgency and fostering efficiency. By combining cognitive load inference with error-type classification, the system selects the most effective adaptive channel (visual, auditory, haptic, or mixed). This ensures trainees remain in an optimal learning zone, where frustration is minimized and skill acquisition is accelerated through targeted, adaptive interventions. 


In addition to cognitive load inference, the system also monitors behavioral performance metrics such as error frequency and repetition. When a user makes the same mistake multiple times despite receiving guidance the system interprets this as procedural confusion and shows ghost hand demonstrations, which provide an embodied model of the required action, showing how to complete the task. By adapting scaffolding to cognitive load, CLAd-VR thus applies Cognitive Load Theory in practice. High load is eased by reducing extraneous demands with ghost hands, arrow cues, and simplified guidance, while low load is challenged through error injection, reflective prompts, and time pressure. These adjustments free capacity for germane load, keeping learners in their optimal zone where engagement is sustained and skills are effectively built.

\section{Limitations and Future Work}
While this work presents the system design for CLAd-VR, we next plan to conduct controlled user studies to evaluate CLAd-VR's usability and training impact. Currently limited by single-modality EEG sensing, a small pilot dataset, and rule-based adaptation, future work will expand data collection, integrate multimodal sensing (e.g., eye tracking, HRV), and develop more theory-driven adaptive strategies. We also aim to extend training beyond CNC drilling and assess long-term effects on skill retention and transfer, enhancing the system’s generalizability and effectiveness.

\bibliographystyle{abbrv-doi}
\balance
\bibliography{references, NSF_SLAI_references}
\end{document}